\documentclass[useAMS,usenatbib,referee]{mn2e}

\usepackage{graphicx}

\title[Precession and Nutation in $\eta$ Carinae]
{Precession and Nutation in the $\eta$ Carinae binary system: Evidences from the X-ray light curve}
\author[Z. Abraham and D. Falceta-Gon\c{c}alves]
{Z. Abraham$^{1}$\thanks{E-mail:zulema@astro.iag.usp.br}, D. Falceta-Gon\c{c}alves$^{2}$ \\
$^{1}$Instituto de Astronomia, Geof\'\i sica e Ci\^encias Atmosf\'ericas, Universidade de S\~ao Paulo, Rua do Mat\~ao 1226,\\ Cidade Universit\'aria 05508-090, S\~ao Paulo, Brazil\\
$^{2}$N\' ucleo de Astrof\' isica Te\' orica, CETEC, Universidade Cruzeiro do Sul, Rua Galv\~ ao Bueno 868, CEP 01506-000 S\~ao Paulo, Brazil\\
\\}

\begin{document}

\date{ }

\pagerange{\pageref{firstpage}--\pageref{lastpage}} \pubyear{2005}

\maketitle

\label{firstpage}

\begin{abstract}

It is believed that $\eta$ Carinae is actually a massive binary system, with the wind-wind interaction  responsible for the strong X-ray emission. 
Although the overall shape of the  X-ray light curve can be explained by the high eccentricity of the binary orbit, other features like the asymmetry near periastron passage and the short quasi-periodic oscillations seen at those epochs, have not yet been accounted for. In this paper we explain these features assuming that the rotation axis of $\eta$ Carinae 
is not perpendicular to the orbital plane of the binary system. 
As a consequence, the companion star will face $\eta$ Carinae on the orbital plane at different latitudes for different orbital phases and, since both the mass loss rate and the wind velocity  are latitude dependent, they would produce the observed asymmetries in the X-ray flux. 
We were able to reproduce the main features of the X-ray light curve assuming that the rotation axis of $\eta$ Carinae forms an angle of $29^\circ \pm 4^\circ$ with the axis of the binary orbit. 
We also explained the short quasi-periodic oscillations by assuming nutation of the rotation axis, with amplitude of about $5^\circ$ and period of about 22 days. The nutation parameters, as well as the precession of the apsis, with a period of about 274 years, are consistent with what is expected from the torques induced by the companion star.

\end{abstract}

\begin{keywords}
stars: individual: $\eta$ Carinae -- stars: binaries: general -- stars: winds 
\end{keywords}
      
\section{Introduction}

The intensity and spectrum of the high energy X-ray flux, and its strict periodicity, are probably the strongest evidence  of the binary nature of the $\eta$ Carinae system. 
The 2-10 keV X-ray emission of $\eta$ Carinae is monitored by the {\it Rossi X-Ray Timing Explorer RXTE} since 
1996, and the published results cover two cycles in the 5.52 year periodic light curve \citep{cor05}. The 
duration of the shallow minima, as well as the general qualitative behavior of the light 
curve, were similar in the two cycles. 
The long lasting intervals of almost stationary intensity were modulated by low amplitude quasi-periodic flares, and the large flux increase that occurred before the minima was enhanced by strong short duration flares \citep{ish99}. 
Although the X-ray light curve was successfully reproduced by analytical approximations involving wind-wind collisions \citep{ish99,cor01} and by numerical simulations \citep{pit98,oka08}, which also reproduced the high resolution spectra obtained with {\it Chandra} \citep{pit02}, some features are still controversial, like the asymmetry near periastron passage, the short quasi-periodic oscillations seen at those epochs, and the  difference in the phases of these oscillations  between the two cycles \citep{oka08,par09}. 

Besides X-rays, other observational features can be related to wind-wind collision. \citet{abr07} were able to reproduce the HeII $\lambda 4686$ line profiles and mean velocities detected close to the 2003.5 minimum by \citet{ste04} and \citet{mar06}, assuming that they were formed in the cooling shocked material flowing along the winds contact surface. More important, to reproduce the line profiles reflected in the Homunculus polar cap (Stahl et al. 2005), they had to assume that the Homunculus  axis is not perpendicular to the orbital plane.

The rotation axis of $\eta$ Carinae probably coincides with the axis of the Homunculus; the shape of the nebula and the measured latitude dependent stellar wind velocity are strong indications  that the rotational velocity is close to its critical value \citep{smi02,dwa02}. 
A consequence of the inclination of the rotation axis relative to the axis of the orbital plane is that the secondary star  faces $\eta$ Carinae at different latitudes as it moves along the orbit, and therefore, the latitude dependent velocity and mass loss rate of the primary's wind will affect the intensity of the X-rays produced in the wind-wind collision region.

Also, the large rotational velocity of $\eta$ Carinae will affect its internal mass distribution, which will depart from spherical symmetry. The torque  induced by the companion star  will result in apsidal motions, as seen in other massive binary systems (e.g. see references in Claret \& Gim\'enez 1993). Finally, the inclination of the rotation axis of $\eta$ Carinae will produce nodding motions, which will further affect the strength of the wind-wind collision and the consequent X-ray intensity.

In this paper we will take all these effects into account and calculate the X-ray light 
curve of $\eta$ Carinae using the analytical approximation derived by \citet{uso92} and the orbital parameters found by \citet{abr05} and \citet{abr07}. We will show that for reasonable values of the precession and nutation periods and amplitudes, it is possible to reproduce the asymmetries in the light curve close to periastron passage and the amplitudes and phases of the short quasi-periodic oscillations for the two binary cycles observed by $RXTE$. 


\section{The X-ray emission model}

We will use the model derived by  \citet{uso92} to calculate, at each point of the orbit, 
the X-ray luminosity  originated in the shock heated gas at both sides of the  contact surface, which depends on the mass loss rates  ($\dot {M_p}$ and $\dot {M_s}$) and wind velocities ($V_p$ and $V_s$) of the primary and secondary stars, respectively, and on the distance $D$ between them.
\citet{pit02b} showed that for the $\eta$ Carinae binary system,  the major contribution to the X-ray flux comes from the 
interaction surface of the secondary wind, because of its higher temperature so that the expression 
derived by \citet{uso92} and valid for adiabatic shocks becomes: 
\begin{equation}
 F_X(\theta _s) = \frac {1.3\times 10^{35}}{4\pi d^2D} \biggl(\frac{\dot {M_s}}{V_s}\biggl)^{3/2} (\dot {M_p}V_p)^{1/2}e^{-\tau(\theta_s)},
 \end{equation}
 \noindent
where $d$ is the distance to $\eta$ Carinae, taken as 2.3 kpc, and $\tau(\theta _s)$ the 
optical depth for X-ray absorption; $\dot {M_p}$ and $\dot {M_s}$ are expressed in units of $10^{-5}$ and $10^{-6}$ $M_\odot$ yr$^{-1}$ respectively, $V_p$ and $V_s$  in units of $10^3$ km s$^{-1}$, and $D$ in units of $10^{13}$ cm; $\theta _s$  is the true anomaly, with $\theta _s = 0$ at periastron. We will not take into account any possible cooling of the very dense shocked gas very close to periastron passage \citep{par09}. 

We will assume that $\dot {M_s}$ and $V_s$ have constant values, although \citet{par09} proposed  a reduced secondary wind velocity near periastron to explain the observed change in the X-ray hardness ratio. On the other hand, we will assume that $\dot {M_p}$ and $V_p$ depend on the latitude $\lambda \,(\theta _s)$ at which the orbital plane intercepts the side of $\eta$ Carinae that faces the secondary star, and can be expressed as  \citep{dwa02}:
\begin{eqnarray}
\dot M(\lambda) & = & {\dot M(90\degr)}\bigl [1-\Omega^2 \cos^2 \lambda\bigr ] , \\
V(\lambda) & = & V(90\degr)\bigl [1-\Omega^2 \cos^2 \lambda \bigr ]^{1/2},
\end{eqnarray}
\noindent
 with $\Omega = \omega /\omega _c$; $\omega$ is the rotation velocity and  $\omega _c = (GM_p/R_p^3)^{1/2}$ its critical value; $G$ is the 
gravitational constant, $M_p$ and $R_p$ are the mass and radius of $\eta$ Carinae, 
respectively. These expressions are valid  when  $\Omega$ is close to unity; they were already used to reproduce the observed wind velocity as a function of latitude in $\eta$ Carinae, 
as well as the shape of the Homunculus nebula \citep{smi02,dwa02}.

By replacing eq. (2) and (3) in (1) we obtain:
\begin{equation}
 F_X(\theta _s) = \frac{G(t)}{D}{\biggl [\frac{\dot {M_p}(\lambda)}{\dot {M_p}(90^\circ)}}\biggr ]^{1/2} \bigg [\frac{{V_p(\lambda)}}{{V_p(90^\circ)}}\biggr ]^{1/2} e^{-\tau_p(\theta_s)},
 \end{equation}
 with
 \begin{equation}
 G(t)= \frac {1.3\times 10^{35}}{4\pi d^2}\biggl(\frac{\dot {M_s}}{V_s}\biggr)^{3/2}\biggl[\dot{M_p}(90^\circ) V_p(90^\circ)\ \biggr]^{1/2}e^{\tau _0},
 \end{equation}
 where we have defined $\tau (\theta _s)= \tau _p(\theta _s) + \tau _0(t)$;
$\tau _p(\theta _s)$ represents the  absorption produced  by the wind of $\eta$ Carinae intercepting the line of sight, and $\tau _0(t)$  is constant or a slowly varying, phase independent function of  time, representing all other sources of absorption.


\begin{figure}
\centering
\includegraphics[width=9cm]{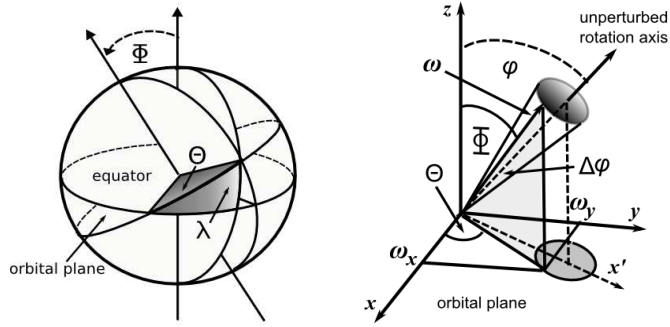}
\caption{{\it Left}: geometrical description of the intersection of the binary system orbital plane with the surface of $\eta$ Carinae at latitude $\lambda$, $\Phi$ is the angle between the rotation axis and the perpendicular to the orbital plane; {\it Right}: definition of the different angles and coordinate systems involved in the precession of the line of apsis and nutation of the 
rotation axis of $\eta$ Carinae}
\label{figure1}
\end{figure}


\subsection{The effects of precession and nutation on  $\lambda(\theta _s)$} 

$\eta$ Carinae must be highly distorted, both by rotation and by the presence of the companion star in a highly eccentric orbit. Although the total angular momentum in a detached binary system is conserved (except for a small amount lost by the stellar winds), energy will be dissipated by the tidal forces, until a minimum energy equilibrium configuration is reached, in which the orbit is circular, and the stellar spins perpendicular to the orbital plane and equal to the orbital velocity. However, although the orbital parameters are continuously changing, the orbit can be consider instantaneously Keplerian, and precession and nutation rates can be calculated as averages over the orbital period \citep{egg98}. 

In this section we will derive the effects of precession and nutation on the latitude $\lambda(\theta _s)$. 
We will assume that $\eta$ Carinae rotates with angular velocity $\vec\omega$ around an axis that  forms an angle $\Phi$ with the perpendicular to the orbital plane ($z$ axis in Figure 1), and its projection on the orbital plane forms an angle $\Theta$ with the line of apsis ($x$ axis), so that:
\begin{equation}
\sin \lambda (\theta _s,t)= \sin \Theta (\theta _s,t)\sin \Phi(t),
\end{equation}
\noindent
where
\begin{equation}
\tan \Theta \,(\theta_s, t)  =  \frac{\omega_y}{\omega_x}, 
\end{equation}
\begin{equation}
\sin \Phi (t)=\frac {\sqrt{\omega_x^2+\omega_y^2}}{\omega}.
\end{equation}
As a consequence of nutation, the rotation angular velocity vector $\vec\omega$ will describe a cone of amplitude $\Delta\varphi$ and period $P_n$ around its non-perturbed direction, which forms an angle $\varphi$ with the polar axis $z$ (Figure 1). 

In a coordinate system $(x^{\prime\prime},y^{\prime\prime},z^{\prime\prime})$, in which  $z^{\prime\prime}$ is directed along the unperturbed rotation axis, the components  of $\vec\omega$ will be $(\omega\sin\Delta\varphi\sin\omega _n(t)$, $\omega\sin\Delta\varphi\cos\omega _n(t)$, $\omega\cos\Delta\varphi )$, with $\omega_n(t)=(2\pi/P_n)\Delta t+\theta _{0n}$; $\theta_{0n}$ is a constant phase and  $\Delta t = t - t_0$, with $t_0 = 2,450,795$ being the JD of the beginning of the 1997.9 minimum. To obtain the components of $\vec\omega$ in the $(x,y,z)$ coordinate system, in which $z$  coincides with the orbital axis, the $(x^{\prime\prime},y^{\prime\prime},z^{\prime\prime})$ axis must be rotated around $y^{\prime\prime}$ by an angle $\varphi$, so that the new $z^\prime$ axis coincides with $z$, and then around  $z^\prime$ by and angle $\theta_ s^\prime =\theta _s + 2\pi/P_p+\theta_{0p}$, to take into account the orbital motion and the precession of the line of apsis; $\theta_{0p}$ is a constant phase.
These rotations can be represented by the matrix $M$:

\begin{eqnarray}
 M = \left(
\begin{array}{ccc}
\cos\theta^\prime_s & \sin\theta^\prime_s & 0 \\
-\sin\theta^\prime_s & \cos\theta^\prime_s & 0 \\
0 & 0 & 1
\end{array}     \right)
   \left(
 \begin{array}{ccc}
\cos\varphi & 0 & \sin\varphi \\
-\sin\varphi &  0 &\cos\varphi \\
0 & 1 & 0
\end{array}     \right),
\end{eqnarray}

\noindent
so that $(x,y,z)^T=M(x^{\prime\prime},y^{\prime\prime},z^{\prime\prime})^T$


\begin{figure*}
\centering
\includegraphics[width=15cm]{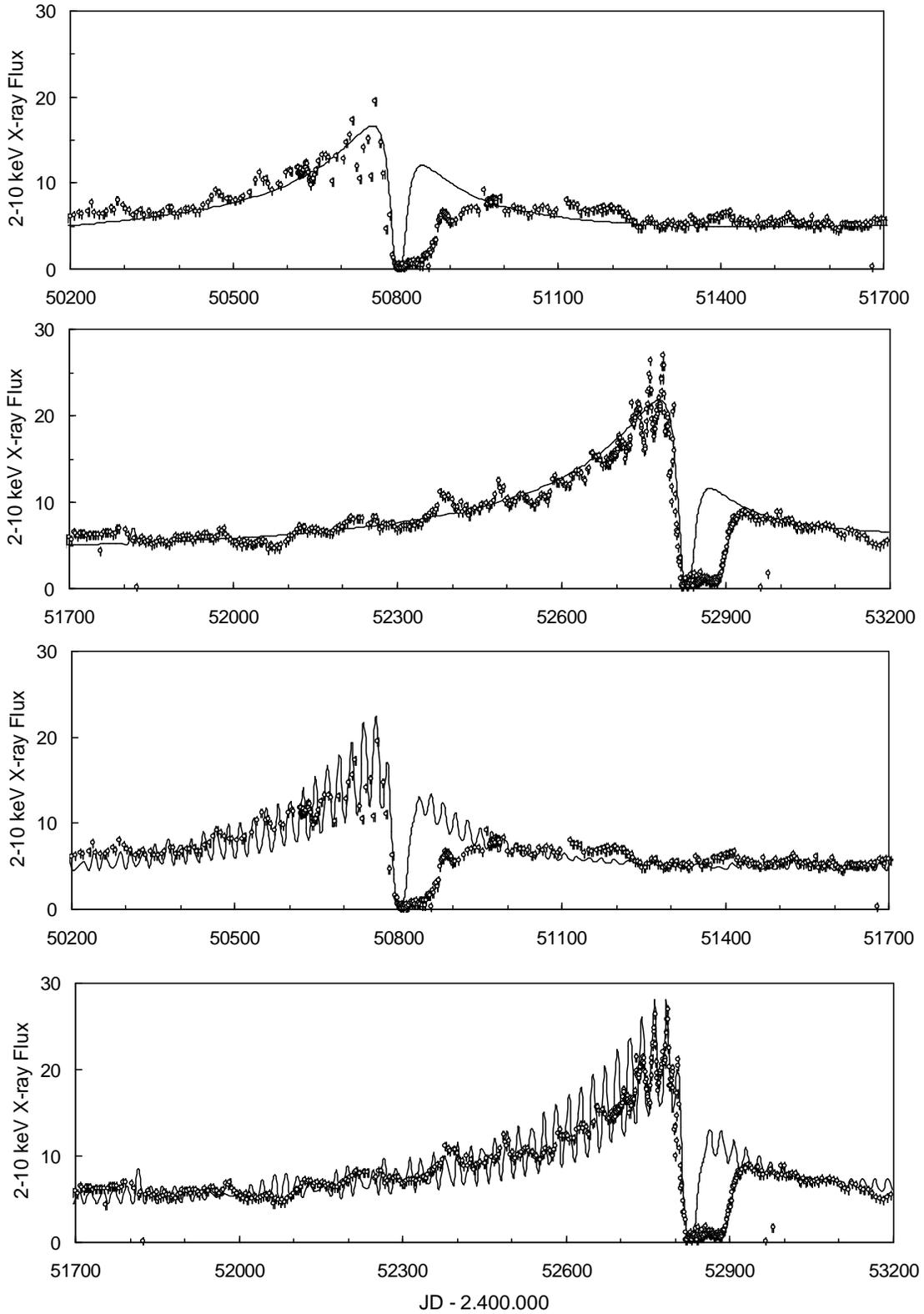}
\caption{Observed 2-10 keV X-ray flux, from Corcoran(2005), shown as open circles, 
and  model, shown as a continuous line, obtained with the parameters listed in Table 1, but with $\Delta\varphi=0^\circ$ (no nutation) in the two upper graphs and $\Delta\varphi=4\fdg5$ for the lower ones.}
\label{figure2}
\end{figure*}


\subsection{Opacity and the  H{\sevensize I} column density}

As mentioned before, the optical depth for X-ray absorption was divided into two parts: $\tau (\theta _s)= \tau _0(t)+\tau _p(\theta _s)$, 
where   $\tau _p(\theta _s)=\sigma_{\rm ph}N_H(\theta_s)$  
represents the photoelectric absorption produced  by the unshocked wind of $\eta$ Carinae intercepting 
the line of sight to the vertex of the X-ray emitting cone, with column density $N_H$, and $\tau _0(t)$ that represents all other sources of absorption, excluding the stellar wind; $\sigma _{\rm ph}$ is 
the cross section of photoelectric absorption, multiplied by the  heavy element abundance relative to H. $N_H(\theta_s)$ is calculated from: 
\begin{equation}
N_H (\theta_s) =\frac{1}{4 \pi \mu m_H}\biggl(\frac{\overline{\dot M_p}}{\overline{V_p}}\biggr)\int^{\infty}_{s_{\rm sh}} \frac {ds}{  s^2+r_0^2} ,
\end{equation}
\noindent
where $\mu$ is the molecular weight and $m_H$ the mass of the hydrogen atom, $\overline{\dot M_p}$ and $\overline{V_p}$ are the mean values of the mass loss rate and wind velocity of the primary star; $s$ is measured along the line of sight to the apex of the X-ray source; $r_0=b\sin \Psi$ and $s_{\rm sh}=b\cos \Psi$, where $b$ is the distance from $\eta$ Carinae to the shock, measured in the orbital plane;  $\Psi$ is calculated from:
\begin{equation}
\sin \Psi = \sin (\theta_s-\theta_0)\sin i ,
\end{equation}
\noindent 
where $i$ is the inclination of the orbit and $\theta_0$ is the true anomaly at conjunction. 
While the inclination is one of our model parameters, the value of the input parameter $\theta_0$ is still controversial \citep{pit98,cor01,fal05,kas07,ham07,abr07,oka08,fal09,par09}. As we will see later, it affects mostly the opacity near periastron, where the model anyway fails to reproduce the duration of the shallow minima. 

By solving the integral of equation (10), we can write:
\begin{equation} 
\tau _{p}(\theta_s) = {\frac {C_\tau} {b\sin \Psi}}\biggl(\frac {\pi}{2}-\arctan {\frac {1}{\tan \Psi}}\biggr),
\end{equation}
\noindent
where $C_\tau=\sigma_{\rm ph}\overline{\dot M_p}/4 \pi \mu m_H \overline{V_p}$.

The contribution to the opacity of the unshocked wind of the secondary secondary star is much smaller than that of the wind of $\eta$ Carinae, because of the much smaller value $\dot M_s/V_s$. However, depending on the position of the secondary star on the orbit near periastron, the absorption due to the shocked gas intercepting the line of sight could be large \citep{fal05,par09} and can affect the duration of the minima in the light curve; this issue will not be addressed here since it requires numerical simulations.  

\subsection{The orbital parameters}

We will use the orbital parameters derived by \citet{abr05} from the observed 7-mm light curve of $\eta$ 
Carinae during the 2003.5 minimum and listed in the left column of Table 1. They were  successfully 
used  to reproduce the  He{\sevensize II}$\lambda 4686$ line profiles and their mean velocities, although to reproduce the mean velocities 
and profiles of the emission lines reflected in the polar cap of the Homunculus,  it was necessary to assume that
the Homunculus axis  is not perpendicular to the orbital plane. Considering  that the angle between this axis 
and the line of sight is $45^\circ $ \citep{dav01,smi06}, for each inclination $i$ of the orbit relative to the 
observer,  an orientation for the Homunculus axis   
$i^*$ was found, for which the reflected line profiles and velocities could be reproduced \citep{abr07}. In Table 2 we present the values of these angles for two values of  $\eta=\dot M_sV_s/\dot M_pV_p$.
In the next section we will use the value of $i^*=\varphi + \Delta \varphi$, obtained from the model that reproduced the observed 
X-ray light curve, to constrain the value of the orbital inclination $i$.


\begin{table}
\centering
\caption{Input and model parameters}
\begin{tabular}{@{}ll}
\hline
Input & Model   \\
\hline
$e\;\;= 0.95$ &  ${P_p} =274 \pm 15$ years\\      
$\theta_0=-45^\circ$ & $\theta_{0p}= 5\fdg3\pm 1^\circ$ \\ 
$P=2024$ days  & $P_n= 22.45\pm 0.04\; $days\\  
$A=15$ A.U. & $\theta_{0n}=90^\circ\pm 10^\circ$\\ 
$t_0= 2,450,795 $  JD & $\Omega= 0.975 \pm 0.001$\\ 
& $\varphi=29^\circ\pm 4^\circ $\\
& $\Delta\varphi\ = 4\fdg5\pm 0\fdg5$\\
& $C_x= (7\pm 1)\times 10^{-5} $\\
& $C_\tau=7.7\pm 0.5$ A.U.\\
& $G_0= (2200\pm 150)\times 10^{-11}$ erg cm$^{-2}$s$^{-1}$\\
& $i = 60^\circ $ \\
\hline
\end{tabular}
\end{table}

\begin{table}
\centering
\caption{Inclination of the binary system orbit $i$ and of $\eta$ Carinae rotation axis $i^*$}
\begin{tabular}{@{}lcc}
\hline
$i$ &  $i^*(\eta=0.2)$ & $i^*(\eta=0.1)$\\
90 & 59 & 57\\
80 & 52 & 47\\
70 & 45 & 40\\
60 & 40 & 33\\
50 & 36 & 25\\
45 & 34 & 22\\
40 & 34 & 23\\
\hline
\end{tabular}
\end{table}


\begin{figure}
\centering
\includegraphics[width=8cm]{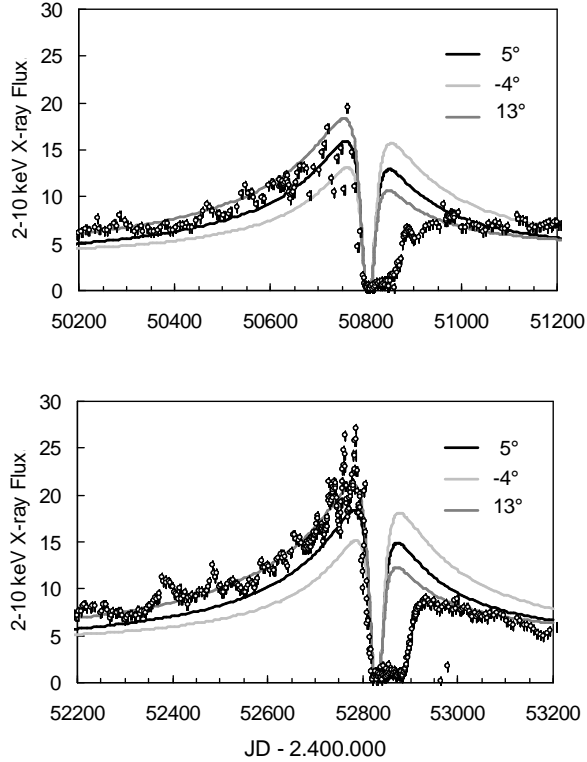}
\caption{Details of the X-ray emission model without precession and nutation, close to the epochs of the two shallow minima,  for three different values for the precession phase $\theta_{0p}$: $-4^\circ, 5^\circ$ and $13^\circ$. All other model parameters are shown in Table 1. }
\label{figure3}
\end{figure}


\begin{figure}
\centering
\includegraphics[width=8cm]{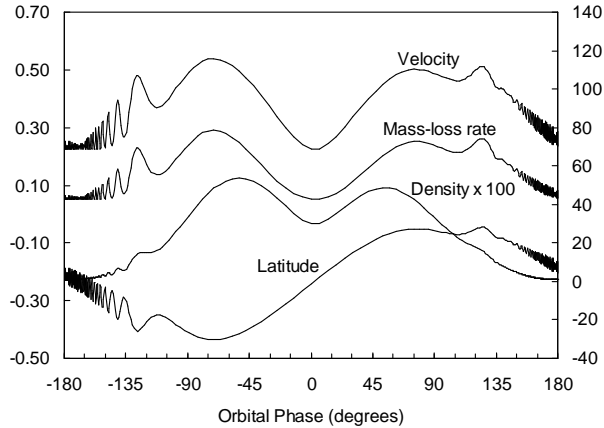}
\caption{Variations with orbital phase of the wind velocity, mass loss rate, wind density at the contact surface, and latitude at the intersection of $\eta$ Carinae with the orbital plane facing the companion star. The first two quantities, given by the left axis,  are relative to their values at $\lambda = 90^\circ$, the latitude values, in degrees, are given in the right axis. The wind density, relative to its value at $\lambda=90^\circ$, in units of $1.5\times 10^{11}$ cm$^{-1}$ is also displayed  at the right axis.}
\label{figure4}
\end{figure}

\section{Results}
We used equations (1) to (12) to model the X-ray light curve of $\eta$ Carinae, with the orbital parameters listed in the first column of Table 1. 
No formal fitting was attempted; instead, the model parameters were changed until they reproduced the general shape of the light curve (except the shallow  minima), and the amplitude and period of the oscillations that occurred just before them. The parameters that fulfilled these criteria are listed in the last  column of Table 1, and the model superposed to the observed light curve is presented in Figure 2; the two upper graphs represent the model without nodding motions ($\Delta\varphi=0$), and the last two graphs include  nodding. The multiplicative term $G(t)$ in equation (5)
was fitted by a function $G(t)=G_0[1+C_x(t-t_0)]$, with $G_0$ and $C_x$ having constant values.  
 The quoted errors were obtained by changing the values of each parameter  while keeping the others constant until the model was no longer acceptable. As mentioned before, no formal fitting was attempted,  but the combination of model parameters could not be changed arbitrarily, since each of them affect some particular feature of the X-ray light curve, as discussed bellow.

The parameters $\Omega$ and $\theta_{0p}$ are responsible for the asymmetry in the light curve at both sides of the shallow minima, as can be seen in the first two graphs in Figure 2. To illustrate this dependence, we plotted in Figure 3 the model light curves close to the minima for $\Omega = 0.975$ and several values of $\theta_{0p}$, when no precession or nutation are present. We found that $\theta_{0p}= 5\fdg3$ reproduces well the asymmetry in the mean value of the light curve close to the minimum of 1997.9, but  $\theta_{0p}\sim 13^\circ$ was necessary to reproduce the minimum of 2003.5. The difference in angles was attributed to the motion of the  apsis, resulting in the precession period listed in Table 1. A large value of $\Omega$ was necessary to get the observed discontinuity; it indicates that $\eta$ Carinae is rotating at almost its critical velocity, as expected from the episodes of large mass loss. Of course, if other causes were responsible for the asymmetry in the X-ray light curve, like changes in opacity or stellar wind parameters, the value of these parameters should be revised.

The nutation parameters $P_n$, $\theta_{0n}$ and $\Delta\varphi$ are responsible for the amplitude and period of the large oscillations  that occur before the minima, as can be seen in the two lower graphs in Figure 2; it is important to notice that, near periastron, the oscillations remained in phase in the two cycles for the same initial phase $\theta_{0n}$. However, the model does not reproduce the amplitude and the period of the oscillations far from periastron, which could be expected from the variation of the torque of the secondary star acting on $\eta$ Carinae. 

The opacity parameter $C_\tau$ determines the shape of the light curve before the minima, and has no influence at other phases.
The linear dependence of $G(t)$ with time implies an increase of 
about 30\% in the X-ray intensity between the two cycles,  which coincides with an
increase in the optical flux during the same time interval \citep{mar04} and could maybe attributed to an overall decrease in opacity.

No assumptions were made on the values of the mass loss rates and wind velocities of the binary stars. Instead, the value obtained for the model parameter $G_0$ and the comparison between the model hydrogen column density $N_H$ and that inferred from the {\it XMM} X-ray spectra observed far from periastron passage \citep{ham07} allowed us to put some constrains  on their magnitudes. From eq. (5) and assuming $\tau_0 = 0$ we can write:

\begin{equation}
\frac{\dot M_s^2}{V_s}= \frac{G_0}{1.83\times 10^{-10}}\eta^{1/2}(90^\circ)
\end{equation}

\noindent
where 

\begin{equation}
\eta(90^\circ)=\frac{\dot M_sV_s}{\dot M_p(90^\circ) V_p(90^\circ)}=
\frac{\dot M_p(\lambda)V_p(\lambda)}{\dot M_p(90^\circ) V_p(90^\circ)}\eta(\lambda)
\end{equation}

From our model $\eta(90^\circ)=0.15\,\eta(30^\circ)$, resulting that for $\eta(30^\circ)=0.2$ and $V_s=3000$ km s$^{-1}$, $\dot M_s=8\times 10^{-6}$ M$_\odot$ y$^{-1}$, well within the values of the mass loss rate of the secondary star used in the literature.

The hydrogen column density inferred from the X-ray spectra in January 2003 was $N_H=9\times 10^{22}$ cm$^{-2}$ \citep{ham07}.
Using this value in equation (10) we obtain $\dot M_p/V_p=4.4\times 10^{14}$ g cm$^{-1}$, where we have used:
\begin{equation}
\int^{\infty}_{s_{\rm sh}} \frac {ds}{  s^2+r_0^2}={\frac {1} {b\sin \Psi}}\biggl(\frac {\pi}{2}-\arctan {\frac {1}{\tan \Psi}}\biggr)=0.062\,\, {\rm AU}^{-1}
\end{equation}
 Assuming a wind velocity of 500 km s$^{-1}$ for $\eta$ Carinae, we find $\dot M_p = 3.3\times 10^{-4}$ M$_\odot$ y$^{-1}$, also consistent with the values found in the literature.

Finally, the model opacity for January 2003 was $\tau_p=0.69$, which together with the observed hydrogen column density gives a value for $\sigma_{ph}=\tau_p/N_H = 4.9$ cm$^2$ g$^{-1}$, consistent with the opacity to 3 keV photons of a 10$^4$-10$^6$ K gas \citep{par08}. 

From the inclination of the rotation axis of $\eta$ Carinae we estimated that, relative to the observer, the orbit has an 
inclination $i \sim 45^\circ-60^\circ$ , for  $\eta$  varying between 0.2 and 0.1, as can be seen in Table 2.
The variation of the wind 
velocity, mass loss rate and wind density  along the orbital period, as well as the 
variation in $\lambda$ are shown in Figure 4 for the first orbital period.


\begin{figure}
\centering
\includegraphics[width=7.5 cm]{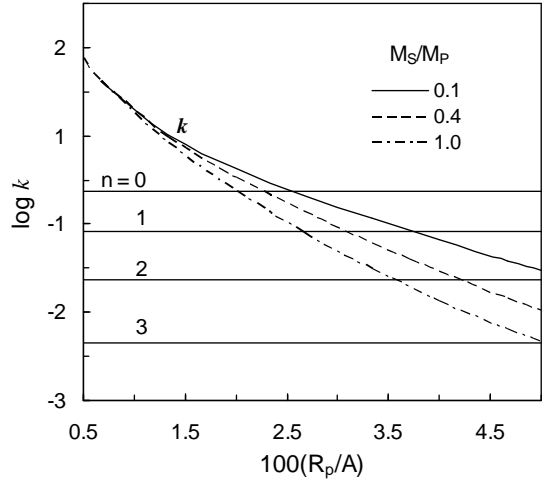}
\caption{Asymmetry parameter $k$ of the internal mass distribution  of $\eta$ Carinae , produced by rotation and gravitational torque induced by the companion star, versus the ratio between the primary star radius and the orbital semi-major axis, for three values of the primary to secondary mass ratio $M_s/M_p$. Horizontal lines are the expected values of $k$ for polytropes of index $n=0$, 1, 2 and 3. }
\label{figure5}
\end{figure}

Although the eccentricity and period of the binary orbit used in the model agree with those used by other authors \citep{pit98, ish99,cor01,pit02,oka08}, the value of $\theta_0$ is still controversial. The model presented in this paper was calculated for $\theta_0 = -45^\circ$; changing its value to $\theta_0 = +45^\circ$ will  not have any effect on the X-ray emission but affect the absorption, mostly  close to periastrom passage, where the model does not anyway reproduce the X-ray light curve; however, using  $\theta_0 = +45^\circ$ will not reproduce the shape and central velocity of the He{\sevensize II}$\lambda 4686$ and Paschen lines \citep{abr07, fal09}.  


\section{Discussion}

As mentioned before, the model does not adjust the amplitude and period of the oscillations in the X-ray light curve far from periastrom passage. 
 From Figure 2 we can also see that it does not reproduce the duration of the shallow minima. 
In fact, neither the analytical models nor the numerical simulations developed up to the present time were able to 
account for the extended minima as a result of X-ray photoelectric absorption by the dense wind of $\eta$ Carinae 
intercepting the line of sight \citep{pit98, ish99,cor01,pit02,ham07}. Possible explanation are the increase in 
the H column density to $\sim 10^{24-26}$ cm$^{-2}$ due to  additional  material 
provided either by a slowly expanding shell of shocked material formed during periastron passage \citep{fal05},  or  by the primary wind itself, which "engulfs" the secondary star  \citep{oka08}, or simply by the suppression of the secondary wind due to accretion of matter from the close primary star \citep{sok05,aka06} . 


\begin{figure}
\centering
\includegraphics[width=7.5 cm]{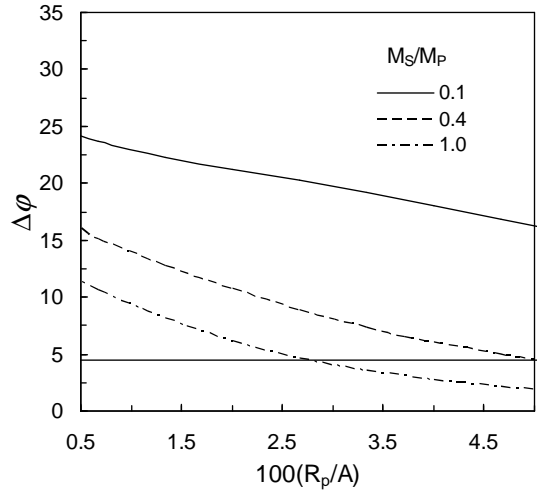}
\caption{Angle $\Delta \varphi$ between the total and orbital angular momenta as a function of $R_p/A$ for several values of the mass ratio $M_s/M_p$. The horizontal line shows the value of $\Delta \varphi$ obtained from the model. }
\label{figure6}
\end{figure}%


\begin{figure}
\centering
\includegraphics[width=7.5 cm]{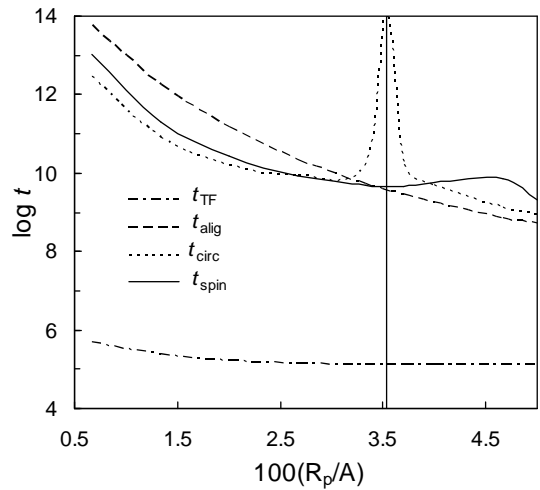}
\caption{Time-scales for tidal friction $t_{TF}$, alignment $t_{alig}$, circularization $t_{circ}$,  and synchronization $t_{spin}$ for $(M_s/M_p=0.4)$. The vertical line separates the regions where the orbit is unstable (left) and stable (right).  }
\label{figure7}
\end{figure}%
\subsection{The orbital stability}
 
 The precession and  nutation parameters are related to the torques of the secondary  acting on the fast rotating non-spherical primary star and can provide constrains on the stellar masses, internal structure and orbital stability. 
\citet{hut82} derived expressions for the evolution of the orbital and spin parameters in  highly eccentric binary system, assuming that the tidal bulges lag by a constant angle from the line that joins the stars. Eggleton et al. (1998) obtained an expression for this angle, considering that the dissipative force is proportional to the rate of change of the quadrupole tensor of the stars, as seen by an observer that rotates with them. 

\begin{figure}
\centering
\includegraphics[width=8 cm]{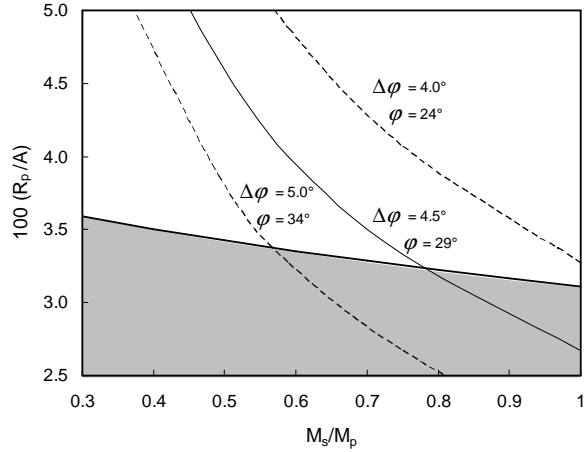}
\caption{Relation between the mass ratio $M_s/M_p$ and $R_p/A$, for the combination of model parameters $(\varphi,\Delta\varphi)=(34^\circ, 5\fdg0)$, $(29^\circ, 4\fdg5)$, and  $(24^\circ, 4\fdg0)$. The shadowed region represents the parameter space in which the orbit is unstable.}
\label{figure8}
\end{figure}%
The precession rate of the apsis, is independent of dissipation and, neglecting the torques of $\eta$ Carinae on the secondary star,  can be expressed as:
\begin{eqnarray}
\frac{P}{P_p} & = & 15\,kf(e)\biggl(\frac{M_s}{M_p}\biggr)\biggl (\frac{R_p}{A}\biggr)^5
\nonumber \\
   & & +k\,\Omega^2 g(e)\frac{(3\cos^2\varphi-1)}{4} \biggl (\frac{R_p}{A}\biggr )^2
\end{eqnarray}
\noindent
with
\begin{eqnarray}
f(e) & = & (1-e^2)^{-5}(1+\frac{3}{2}e^2+\frac{1}{8}e^4)\\
g(e) & = & (1-e^2)^{-2},
\end{eqnarray}
\noindent
where $k$ is the constant part of the quadrupole moment, which depends on the internal mass distribution of the primary star, $A$ is the  semi-major axis of the orbit, and $R_p$ is the radius of the primary star.

In massive binary systems, the measured precession period $P_p$, together with the orbital parameters, stellar masses and radii, are used to calculate $k$ and improve the stellar structure models \citep{clar93}. 
Since for $\eta$ Carinae the orbital and stellar parameters are unknown,  we will take $M_s/M_p$ and $R_p/A$ as free parameters.

In Figure 5 we show the values of $k$  that satisfy  equation (13) for  $P/P_p=0.02$ (obtained from our model) as a function of $R_p/A$, for several values of $M_s/M_p$. The maximum value allowed for $R_p/A$ corresponds to the separation between the stars at periastron: $R_p/A=(1-e)=0.05$. The horizontal lines represent  the values of $k$ for rotating  polytropes with indices $n=0, 1, 2$ and 3 \citep{cha33}; $n=0$ corresponds to a star with constant density while for $n=1$  the radius is independent of the central density.


Another constrain for the ratios $M_s/M_p$ and $R_p/A$ can be obtained from the amplitude of nutation $\Delta\varphi$, which represents the angle between the orbital and total angular momenta, and can be expressed as:
\begin{equation}
\cot \Delta\varphi=\cot \varphi+\frac{h}{I\omega \sin\varphi},
\end{equation}

\noindent
where $h$ is the orbital angular momentum, $I=M_pr_g^2R_p^2$  the momentum of inertia of the primary star and $r_g^2$ a parameter that depends on its internal structure. We can also write $h/I\omega$ in terms of  $M_s/M_p$ and $R_p/A$:

\begin{equation}
\frac{h}{I\omega}=\frac{(1+M_s/M_p)^{1/2}}{M_s/M_p}\biggl(\frac{R_p}{A}
\biggr)^{1/2}\frac{1}{r_g^2\Omega(1-e^2)^{1/2}},
\end{equation}

In Figure 6 we present the relation between $\Delta\varphi$ and $R_p/A$ for $\varphi=29^\circ$ (obtained from our model), and $M_s/M_p=0.1$, 0.4, and 1.0. 
In the calculation, we used an interpolated  relation between $r_g^2$ and $k$ for rotating polytropes obtained from \citet{mot52}:
\begin{equation}
\log r_g^2=0.453\log k -0.307
\end{equation} 
The value of $\Delta \varphi=4.5$ derived by fitting our model to the X-ray data is also shown in the figure. We can see that  $M_s/M_p\sim0.4$ represents the minimum value of the mass ratio for which a solution can be found. However, this result  depends of the actual  value of $r_g^2$,  and should  be consider only in the context of a consistency test for the parameters derived from the X-ray light curve. 

The last parameter derived from the observations is the nutation period, which according to Eggleton et al. (1998) depends on both  the conservative torques and those produced by tidal dissipation and can be written as:

\begin{eqnarray}
\frac{P}{P_n} & = & \frac{k}{2r_g^2}\frac{M_s}{M_p}\biggl(1+\frac{M_s}{M_p}\biggr)^{-1/2}\biggl(\frac{R_p}{A}\biggr)^{3/2}
\frac{\cos\varphi}{(1-e^2)^{3/2}} \\
 & + &\frac{3}{8r_g^2}\frac{M_s}{M_p}\biggl(1+\frac{M_s}{M_p}\biggr)^{-1}\biggl(\frac{A}{R_p}\biggr)^2
\frac{e^2(1+1/6e^2)}{(1-e^2)^{9/2}}\frac{P}{2\pi t_{TF}}\nonumber,
\end{eqnarray}

\noindent
where  $t_{TF}$ is the tidal friction time scale.

When only non-dissipative torques, represented by the first tem in eq. (19), are considered, the nutation period turns out to be several orders of magnitude larger than that obtained from our model $(P_n\cong P/92)$ for any combination of $k$, $(R_p/A)$ and $(M_s/M_p)$ given by equation (13); therefore, the observed nutation must be produced by the dissipative torques, i.e. the second term in eq. (19), and its value can be used to estimate the dissipative time-scale $t_{TF}$:
\begin{equation}
t_{TF}=\frac{3}{8r_g^2}\frac{M_s}{M_p}\biggl(1+\frac{M_s}{M_p}\biggr)^{-1}\biggl(\frac{A}{R_p}\biggr)^2
\frac{e^2(1+1/6e^2)}{(1-e^2)^{9/2}}\frac{P_n}{2\pi}.
\end{equation}
In Figure 7 we display $t_{TF}$ as a function of $R_p/A$ for $M_s/M_p =0.4$.

When the tidal dissipation time-scale is known, it is possible to determine the time-scales for spin alignment and synchronization, and orbit circularization, defined as:
\begin{equation}
 \frac {y}{t_y}=\left| \frac {dy}{dt}\right |,
 \end{equation}
 
 \noindent
 where $t_y = ( t_{alig}, t_{spin}, t_{circ})$ for $y= (\varphi, \omega, e)$, respectively. 

For the case of a highly eccentric orbit, they can be obtained from \citet{hut82}:

\begin{equation}
t_{alig}=t_{TF}\biggl[\frac{693}{16}\frac{k}{r_g^2}\biggl(\frac{M_s}{M_p}\biggr)^2\biggl(\frac{R_p}{A}\biggr)^6
\frac{H_1}{(1-e^2)^6}\biggr]^{-1},
\end{equation}
\begin{equation}
t_{spin}=t_{TF}\biggl[\frac{693}{16}\frac{k}{r_g^2}\biggl(\frac{M_s}{M_p}\biggr)^2\biggl(\frac{R_p}{A}\biggr)^6
\frac{H_2}{(1-e^2)^6}\biggr]^{-1},
\end{equation}
\begin{equation}
t_{circ}=t_{TF}\biggl[\frac{11583}{64}k\biggl(1+\frac{M_s}{M_p}\biggr)\biggl(\frac{R_p}{A}\biggr)^8
\frac{H_3}{(1-e^2)^{13/2}}\biggr]^{-1},
\end{equation}
\noindent
with
\begin{eqnarray}
H_1 & = & \frac{h_1(\varepsilon)}{\Omega}\biggl(1+\frac{M_s}{M_p}\biggr)^{1/2}\biggl(\frac{R_p}{A}\biggr)^{3/2}
\nonumber \\
& - & \frac{20}{33}\frac{(1-e^2)^{3/2}}{(1+e)^2}(1-\eta)h_2(\epsilon),
\end{eqnarray}
\begin{eqnarray}
H_2 & = & \frac{h_1(\varepsilon)}{\Omega}\biggl(1+\frac{M_s}{M_p}\biggr)\biggl(\frac{R_p}{A}\biggr)^{3/2}
\nonumber \\
& - & \frac{40}{33}\frac{(1-e^2)^{3/2}}{(1+e)^2}h_2(\epsilon),
\end{eqnarray}
\begin{eqnarray}
H_3  & = & h_3(\varepsilon)
\\
& - & h_4(\varepsilon)\Omega\frac{112}{117}\frac{(1-e^2)^{3/2}}{(1+e)^2}\biggl(1+\frac{M_s}{M_p}
\biggr)^{-1/2}\biggl(\frac{R_p}{A}\biggr)^{-3/2}
\nonumber 
\end{eqnarray}
\begin{equation}
\eta=\frac{I\omega}{h},
\end{equation}
\begin{equation}
\epsilon=1-e
\end{equation}
\begin{equation}
h_1(\varepsilon)=1-\frac{30}{11}\varepsilon+\frac{35}{11}\varepsilon^2-\frac{460}{231}\varepsilon^3+
\frac{5}{7}\varepsilon^4-\frac{10}{77}\varepsilon^5\frac{5}{231}\varepsilon^6,
\end{equation}
\begin{equation}
h_2(\varepsilon)=1-\frac{19}{7}\varepsilon+\frac{443}{140}\varepsilon^2-\frac{69}{35}\varepsilon^3+
\frac{51}{70}\varepsilon^4-\frac{6}{35}\varepsilon^5\frac{3}{140}\varepsilon^6.
\end{equation}
\begin{equation}
h_3(\varepsilon)=1-\frac{30}{13}\varepsilon+\frac{345}{143}\varepsilon^2-\frac{580}{429}\varepsilon^3+
\frac{5}{11}\varepsilon^4-\frac{10}{143}\varepsilon^5\frac{5}{429}\varepsilon^6,
\end{equation}
\begin{equation}
h_4(\varepsilon)=1-\frac{7}{3}\varepsilon+\frac{205}{84}\varepsilon^2-\frac{29}{21}\varepsilon^3+
\frac{19}{42}\varepsilon^4-\frac{2}{21}\varepsilon^5\frac{1}{84}\varepsilon^6.
\end{equation}

In Figure 7 we plotted $t_{alig}, t_{spin}$, and $t_{circ}$ as a function of $R_p/A$ for $M_s/M_p=0.4$.
We can see that although $T_{TF}$ is of the order of $10^5$ years, the timescales for alignment, synchronization and circularization are larger than $10^9$ years, which means that the binary system did not reached yet its equilibrium configuration, considering an evolution timescale of $10^6-10^7$ years for the massive stars.

We should notice in eq. (27) that $H_3$ can be negative, implying that the eccentricity can increase  with time, which corresponds to an unstable orbit. 
This occurs when the spin of the primary star is larger than the orbital angular velocity at periastron, or:
\begin{equation}
\frac{R_p}{A}<\frac{1-e^2}{(1+e)^{3/2}}\frac{1}{(1+M_s/M_p)^{1/3}}
\end{equation}

For $M_s/M_p = 0.4$ this occurs when $R_p/A=0.035$, this value is shown as a vertical line in  fig. (7). The general relation given by eq. (34)
is shown in Figure 8; the shadowed area represents the parameter space for which the orbit is unstable. In the same Figure we show the lines that satisfy eq. (16), for the parameters found in our model $(\varphi=29^\circ,\Delta\varphi=4\fdg5)$, and for their extremes given by the uncertainties $(\varphi=34^\circ,\Delta\varphi=5\fdg0)$ and $(\varphi=24^\circ,\Delta\varphi=4\fdg0)$. The values of $M_s/M_p$ and $R_p/A$ for the binary system should lie between these two extreme lines.

\section{Conclusions}
We were able to reproduce the general features of the 2-10 keV X-ray light curve of $\eta$ Carinae obtained by 
$RXTE$ \citep{cor05} during  two cycles, including the amplitudes and phases of 
the short period oscillations that occurred prior to the shallow minima, assuming that the star rotates around an axis that is not perpendicular to the orbital plane, so that the secondary star faces $\eta$ Carinae at different latitudes as it moves along the orbit, using the fact that both the mass loss rate 
and the terminal wind velocity of $\eta$ Carinae are latitude dependent. According to the model, the star should be rotating with a fraction $\Omega =0.975$ of its critical velocity around an axis that forms an angle of about 
$29^\circ$ with the axis of the orbital plane, nutates with an amplitude of about $5^\circ$ and a period of 22.5 days. We also found that the line of apsis precesses  with a period of about 274 years.  
According to the results of \citet{abr07}, the inclination obtained for the rotation axis of $\eta$ Carinae  implies  that the inclination of the binary orbit relative to the observer must be $i\sim 45^\circ-60^\circ$, depending on the ratio of the wind momenta ($\eta \sim 0.2-0.1$).
To fit the overall amplitude of the X-ray light curve, we calculated the phase and latitude dependent opacity due to the wind of 
$\eta$ Carinae intercepting the line of sight, and introduced a phase independent 
absorption, which decreased linearly with time and explained the overall increase in the X-ray flux between the two cycles. 
We used the precession period and nutation amplitude and rate to constrain the mass ratio of the binary system and the radius of the primary star relative to the semi-major orbital axis. We found that the orbit is stable if the radius of $\eta$ Carinae is larger than 0.035 times the orbit major axis, and the mass of the companion star at least half the mass of $\eta$ Carinae. Finally we found that for stable orbits, the time scale for orbit circularization, spin alignment and synchronization is much larger than the lifetime of the stars.

\section*{Acknowledgments}

This work was partially  supported by the Brazilian agencies  
FAPESP and CNPq.

\end{document}